\documentclass[twoside]{LCWS11}
\usepackage[latin1]{inputenc}
\usepackage[dvips]{graphicx,epsfig,color}
\usepackage{wrapfig,rotating}
\usepackage{amssymb,amsmath,array}
\usepackage{cite}
\usepackage{axodraw}
\newcommand{\eqFeyn}[1]{%
\begin{array}{c} #1 \end{array}
}
\newcommand{\be}{\beta}
\newcommand{\mD}{m_D}
\newcommand{\mM}{m_M}

\newcommand{\mR}{m_{\tilde R}}

\newcommand{\Anu}{A_\nu}

\newcommand{\Bnu}{B_\nu}
\newcommand{\mnu}{m_\nu}
\newcommand{\Ynu}{Y_\nu}
\def\hSi{\hat{\Sigma}}
\def\Si{\Sigma}

\newcommand{\DRbar}{\ensuremath{\overline{\mathrm{DR}}}}
\newcommand{\mDRbar}{\ensuremath{\mathrm{m}\overline{\mathrm{DR}}}}

\newcommand{\cp}{{\cal CP}}
\newcommand{\cO}{{\cal O}}
\newcommand{\wz}{\sqrt{2}}
\newcommand{\edz}{\frac{1}{2}}

\newcommand{\fa}{{\em FeynArts}}
\newcommand{\fc}{{\em FormCalc}}
\newcommand{\fh}{{\tt FeynHiggs}}
\newcommand{\MW}{M_W}
\newcommand{\MZ}{M_Z}
\newcommand{\MA}{M_A}

\newcommand{\Mh}{M_h}
\newcommand{\MH}{M_H}

\newcommand{\Snu}{\tilde \nu}

\newcommand{\se}[1]{\Sigma_{#1}}
\newcommand{\ser}[1]{\hat{\Sigma}_{#1}}
\newcommand{\KL}{\left(}
\newcommand{\KR}{\right)}

\newcommand{\tb}{\tan \beta}

\newcommand{\CTb}{\cot \beta}

\newcommand{\CZb}{\cos 2\beta}

\newcommand{\gev}{\,\, \mathrm{GeV}}
\newcommand{\mev}{\,\, \mathrm{MeV}}

\newcommand{\non}{\nonumber}
\newcommand{\id}{{\rm 1\kern-.12em
\rule{0.3pt}{1.5ex}\raisebox{0.0ex}{\rule{0.1em}{0.3pt}}}}
\newcommand{\lsim}
{\;\raisebox{-.3em}{$\stackrel{\displaystyle <}{\sim}$}\;}
\newcommand{\gsim}
{\;\raisebox{-.3em}{$\stackrel{\displaystyle >}{\sim}$}\;}

\newcommand{\tadH}{T_H}
\newcommand{\tadh}{T_h}
\newcommand{\tanb}{\tan \beta\,}

\newcommand{\dmhsq}{\delta m_h^2}

\newcommand{\dZ}[1]{\delta Z_{#1}}

\begin{document}
\title{$\bf  \Mh$ in the MSSM-seesaw scenario with ILC precision}
\author{S.~HEINEMEYER$^{1}$,M.J.~HERRERO$^{2}$, S.~PE\~NARANDA$^{3,4}$, 
A.M.~RODR\'IGUEZ-S\'ANCHEZ$^{2}$
\thanks{Preprint number: IFT-UAM/CSIC-12-09.}
\vspace{.3cm}\\
1- Instituto de F\'isica de Cantabria (CSIC-UC), Santander, Spain\\
\vspace{.1cm}\\
2- Departamento de F\'isica Te\'orica and Instituto de F\'isica Te\'orica,
UAM/CSIC\\
Universidad Aut\'onoma de Madrid, Cantoblanco, Madrid, Spain \\
\vspace{.1cm}\\
3- Departamento de F\'isica Te\'orica, Universidad de Zaragoza, 
Zaragoza, Spain\\
\vspace{.1cm}\\
4- Departament de F{\'\i}sica Fonamental, Universitat de Barcelona, Spain
}

\maketitle

\begin{abstract}
We review the computation of the one-loop radiative corrections from the 
neutrino/ sneutrino sector to the 
lightest Higgs boson mass, $\Mh$, within the context of the so-called 
MSSM-seesaw scenario. This model introduces right handed neutrinos and their 
supersymmetric partners, the sneutrinos, including Majorana mass terms.
We find negative and sizeable corrections to
$\Mh$, up to $-5 \gev$ for a large Majorana scale, $10^{13}-10^{15} \gev$, 
and for the lightest neutrino mass in a range $0.1-1$~eV. 
The corrections to $\Mh$
are substantially larger than the anticipated ILC precision for 
large regions of the MSSM-seesaw parameter space.
\end{abstract}

\section{Introduction}
The current  experimental data 
on neutrino mass differences and 
neutrino mixing angles clearly indicate new physics beyond
the so far successful Standard Model (SM) of Particle Physics.
In particular, neutrino oscillations imply that at
least two generations of neutrinos must be massive. Therefore, 
one needs to extend the SM to incorporate neutrino mass terms.

We have explored~\cite{Heinemeyer:2010eg} the simplest version of a
Supersymmetric extension of the SM,
the well known Minimal Supersymmetric Standard Model (MSSM), extended 
by right-handed Majorana neutrinos and
where the seesaw mechanism of type I~\cite{seesaw:I} is implemented to 
generate the small neutrino masses. For simplicity, as a first step, 
we focus here in the one generation case. 

On the other hand, it is well known that heavy Majorana neutrinos, 
with a Majorana mass scale $m_M \sim 10^{13}-10^{15} \gev$, 
induce large LFV rates~\cite{LFV}, due to their potentially large Yukawa
couplings to the Higgs sector. For the same reason, radiative corrections to
Higgs boson masses due to such heavy Majorana neutrinos 
could also be relevant. 
Consequently, our study has been  focused on the 
radiative corrections to the lightest MSSM $\cp$-even $h$ boson mass, $\Mh$,
due to the one-loop contributions from the neutrino/sneutrino sector 
within the MSSM-seesaw framework. 

In the following we briefly review the main relevant aspects of 
the calculation of the mass corrections and the numerical results. 
Further details can be found in~\cite{Heinemeyer:2010eg},
where also an extensive list with references to previous works can be found.

\section{The MSSM-seesaw model}
\label{sec:nusnu}

The MSSM-seesaw model with one neutrino/sneutrino generation is described
in terms of the well known MSSM superpotential plus the new
relevant terms given as: 
\begin{equation}
\label{W:Hl:def}
W\,=\,\epsilon_{ij}\left[\Ynu \hat H_2^i\, \hat L^j \hat N \,-\, 
Y_l \hat H_1^i\,\hat L^j\, \hat R  \right]\,+\,
\edz\,\hat N \,\mM\,\hat N \,,
\end{equation}
where $\hat N = (\Snu_R^*, (\nu_R)^c)$ is 
the additional superfield that contains the  right-handed 
neutrino $\nu_{R}$ and its scalar partner $\Snu_{R}$.   

There are also new relevant terms in the soft SUSY breaking potential: 
\begin{equation}
V^{\Snu}_{\rm soft}= m^2_{\tilde L} \Snu_L^* \Snu_L +
  m^2_{\tilde R} \Snu_R^* \Snu_R + (\Ynu \Anu H^2_2 
  \Snu_L \Snu_R^* + \mM \Bnu \Snu_R \Snu_R + {\rm h.c.})~.
\end{equation}

After electro-weak (EW) symmetry breaking, the charged lepton and 
Dirac neutrino masses
can be written as
\begin{equation}
m_l\,=\,Y_l\,\,v_1\,, \quad \quad
\mD\,=\,\Ynu\,v_2\,,
\end{equation}
where $v_i$ are the vacuum expectation values (VEVs) of the neutral Higgs
scalars, with $v_{1(2)}= \,v\,\cos (\sin) \be$ and $v \simeq 174 \gev$.

The $ 2 \times 2$ neutrino mass matrix is given in terms of $\mD$ and
$\mM$ by: 
\begin{equation}
\label{seesaw:def}
M^\nu\,=\,\left(
\begin{array}{cc}
0 & \mD \\
\mD & \mM
\end{array} \right)\,. 
\end{equation}
Diagonalization of $M^\nu$ leads to two mass
eigenstates, which are Majorana fermions
with the respective  mass eigenvalues given by:
\begin{equation}
\label{nuigenvalues}
m_{\nu,\, N}  = \edz \KL \mM \mp \sqrt{\mM^2+4 \mD^2} \KR~. 
\end{equation}  
In the seesaw limit, i.e. when $\xi \equiv \frac{\mD}{\mM} \ll 1~$, one finds,
\begin{align}
m_{\nu}= -\mD \xi + \mathcal{O}(\mD \xi^3) \simeq -\frac{\mD^2}{\mM} ~,\quad \quad m_N    &=  \mM + \mathcal{O}(\mD \xi) \simeq \mM ~.
\end{align}
   
Regarding the sneutrino sector, the sneutrino mass matrices for the
$\cp$-even, ${\tilde M}_{+}$,  and the $\cp$-odd, ${\tilde M}_{-}$,
subsectors are given respectively by 
\begin{equation}
{\tilde M}_{\pm}^2=
\left( 
\begin{array}{cc} m_{\tilde{L}}^2 + \mD^2 + \edz \MZ^2 \cos 2 \be & 
\mD (A_{\nu}- \mu \CTb \pm \mM) \\  
\mD (A_{\nu}- \mu \CTb \pm \mM) &
m_{\tilde{R}}^2+\mD^2+\mM^2 \pm 2 \Bnu \mM \end{array} 
\right)~.
\end{equation}
The diagonalization of these two matrices, ${\tilde M}_{\pm}^2$, 
leads to four sneutrino mass eigenstates. In the seesaw limit,
 where $\mM$ is much bigger than all the
other scales the corresponding sneutrino masses are given by: 
\begin{eqnarray}
 m_{{\Snu_+},{\Snu_-}}^2
&=& m_{\tilde{L}}^2 + 
\edz \MZ^2 \CZb \mp 2 \mD (A_{\nu} -\mu \CTb-\Bnu)\xi ~, \non \\
 m_{{\tilde N_+},{\tilde N_-}}^2  &=& \mM^{2} \pm 2 \Bnu \mM + \mR^2 + 2 \mD^2 ~.
\end{eqnarray}  
Finally, in the interaction Lagrangian that is relevant 
for the present work, there are terms already present in the MSSM: 
the pure gauge interactions between the
left-handed neutrinos and the $Z$~boson, 
those between the 'left-handed' sneutrinos and the Higgs bosons, and those 
between the 'left-handed' sneutrinos and the $Z$~bosons. 
In addition, in this MSSM-seesaw scenario, there are interactions 
driven by the neutrino Yukawa couplings (or equivalently $\mD$ since 
$\Ynu=(g\mD)/(\wz \MW \sin\be)$), 
as for instance $g_{h\nu_L \nu_R} =  -\frac{igm_D \cos\alpha}{2M_W\sin\beta}$, 
and new interactions due to the
Majorana nature driven by $\mM$, which are not present in the case of Dirac
fermions, as for instance 
$g_{h\Snu_L\Snu_R}=-\frac{igm_Dm_M\cos\alpha}{2M_W\sin\beta}$. 
Besides, the Higgs boson sector in the MSSM-seesaw model is as in the MSSM.

\section{Calculation}

  In the Feynman diagrammatic (FD) approach the higher-order corrected 
$\cp$-even Higgs boson masses in the MSSM, denoted here as $\Mh$ and $\MH$, 
are derived by finding the 
poles of the $(h,H)$-propagator 
matrix, which is equivalent to solving the following equation \cite{mhcMSSMlong}: 
\begin{equation}
\left[p^2 - m_{h}^2 + \hSi_{hh}(p^2) \right]
\left[p^2 - m_{H}^2 + \hSi_{HH}(p^2) \right] -
\left[\hSi_{hH}(p^2)\right]^2 = 0~,
\label{eq:proppole}
\end{equation}
where $m_{h,H}$ are the tree level masses.
The one loop renormalized self-energies, $\hSi_{\phi\phi}(p^2)$,  in \eqref{eq:proppole} can be expressed
in terms of the bare self-energies, $\Si_{\phi\phi}(p^2)$, the field
renormalization constants $\delta Z_{\phi\phi}$  and the mass counter terms  $\delta m_{\phi}^2$, where $\phi$ stands for
$h,H$.
 For example, the lightest Higgs boson renormalized self energy reads:
\begin{equation}\label{rMSSM:renses_higgssector}
\ser{hh}(p^2)  = \se{hh}(p^2) + \dZ{hh} (p^2-m_h^2) - \dmhsq~.
\end{equation}

Regarding the renormalization prescription, 
we have used an on-shell renormalization scheme for 
$\MZ, \MW$ and $\MA$ mass counterterms 
 and $\tadh, \tadH$  tadpole counterterms. 
On the other hand, we have used a modified $\DRbar$ scheme  
$\left(\mDRbar\right)$ for the renormalization of the wave function 
and $\tanb$. The m$\DRbar$ scheme  is very similar to
the well known $\DRbar$ scheme but instead of subtracting the usual 
term proportional to 
$\Delta= \frac{2}{\epsilon}-\gamma_E+ \log(4 \pi)$ one subtracts
the term proportional to 
$\Delta_m = \Delta -\log(m^2_M/\mu^2_{\DRbar})$, hence, 
avoiding large logarithms of the large scale $m_M$.
As studied in other works~\cite{decoup1}, this scheme minimizes 
higher order corrections when two very different scales are involved 
in a calculation of radiative corrections.

The full one-loop  $\nu/\Snu$ corrections to the self-energies,
$\hSi_{hh}^{\nu/\Snu}$, $\hSi_{HH}^{\nu/\Snu}$ and $\hSi_{hH}^{\nu/\Snu}$,
 entering~(\ref{eq:proppole}) have been evaluated with  
\fa $\,$ and \fc~\cite{FAFC}. The new Feynman rules 
for the $\nu/\Snu$ sector are inserted into a new model file. 
Since we are interested in exploring the relevance of the new 
radiative corrections to $\Mh$ from the neutrino/sneutrino sector, 
we will present here our results in terms of the mass
difference with respect to the MSSM prediction. Consequently, we define,
\begin{equation}
\Delta m_h^{\mDRbar} := \Mh^{\nu/\Snu} - \Mh, 
\end{equation}
where $\Mh^{\nu/\Snu}$ denotes the pole for the 
light Higgs mass including the $\nu/\Snu$
corrections (i.e.\ in the MSSM-seesaw model), and $\Mh$ the 
corresponding pole in the MSSM, i.e without the $\nu/\Snu$
corrections. Thus, for a given set of input parameters we first 
calculate $\Mh$ in the MSSM with the
help of \fh~\cite{feynhiggs}, such that all relevant known 
higher-order corrections are included. Then we add the new contributions from
the neutrino/sneutrino sector and eventually compute 
$\Delta m_h^{\mDRbar}$.

\section{Results}

We have obtained the full analytical results for the renormalized Higgs 
boson self-energies and their expressions in the seesaw limit. 
In order to understand in simple terms the 
analytical behavior of our full numerical results we have expanded 
the renormalized self-energies in
powers of the seesaw parameter $\xi=\mD/\mM$:
\begin{equation}
\hSi(p^2)=\left(\hSi(p^2)\right)_{\mD^0}+\left(\hSi(p^2)\right)_{\mD^2}+
\left(\hSi(p^2)\right)_{\mD^4} + \ldots ~.
\label{seesawser}
\end{equation}
The zeroth order of this expansion is precisely the pure gauge contribution 
and it does not depend on $\mD$ or $\mM$. Therefore, it corresponds to the 
result in the MSSM. The rest of the terms of the expansion are 
the Yukawa contribution. The leading term of this Yukawa contribution 
is the ${\cal O}(m_D^2)$ term, because 
it is the only one not suppressed by the Majorana scale. 
In fact it goes as $Y_\nu^2M^2_{\rm EW}$, where  $M^2_{\rm EW}$ denotes 
generically the electroweak scales involved, 
concretely, $p^2$, $M_Z^2$ and $M_A^2$. In particular, the ${\cal O}(p^2m_D^2)$  terms of the renormalized self-energy,
 which turn out to be the most relevant leading contributions, 
separated into the neutrino and sneutrino contributions, read:

\begin{equation}
\left.\hSi_{hh}^{\mDRbar}\right|_{\mD^2 p^2} \sim
\left.
\eqFeyn{%
\begin{picture}(70,40)
\DashLine(10,25)(25,25){3}
\Text(8,25)[r]{$h$}
\DashLine(55,25)(70,25){3}
\Text(72,25)[l]{$h\,+$}
\ArrowArc(40,25)(15,0,180)
\Text(40,46)[b]{$\nu_L$}
\ArrowArc(40,25)(15,-180,0)
\Text(40,4)[t]{$\nu_R$}
\end{picture}
}
\mbox{~~}
\eqFeyn{%
\begin{picture}(70,40)
\DashLine(10,25)(25,25){3}
\Text(8,25)[r]{$h$}
\DashLine(55,25)(70,25){3}
\Text(72,25)[l]{$h$}
\DashArrowArc(40,25)(15,0,180){3}
\Text(40,46)[b]{$\Snu_L$}
\DashArrowArc(40,25)(15,-180,0){3}
\Text(40,4)[t]{$\Snu_R$}
\end{picture}
}\,\,\right|_{\mD^2 p^2}
\sim \frac{g^2 p^2 \mD^2 c^2_{\alpha}}{64 \pi^2  \MW^2 s^2_{\beta} } + \frac{g^2 p^2 \mD^2 c^2_{\alpha}}{64 \pi^2  \MW^2 s^2_{\beta} }.
\end{equation}
Notice that the above neutrino contributions come from the Yukawa interaction
 $g_{h\nu_L \nu_R}$, which is extremely suppressed in the Dirac case but can
be large in the Majorana case. The sneutrino contributions come from
the new couplings $g_{h\Snu_L\Snu_R}$, which are not present in the Dirac case. 
It is also interesting to remark  that these terms, being $\sim p^2$, depend
on the external momentum. Therefore, 
at large $\mM$, to keep just the Yukawa part is
a good approximation, but to neglect the momentum dependence or to set 
the external momentum to zero are certainly not.
In consequence, the effective potential method will not provide a
realistic result for the radiative corrections to the Higgs mass.
Similarly, obtaining the leading logarithmic terms 
in a RGE computation, would also miss these finite terms.

The behaviour of the renormalized self-energy with all others parameters
entering in the computation have been 
discussed in~\cite{Heinemeyer:2010eg}.     
According to our detailed analysis in this paper, 
the most relevant parameters for our purposes 
are: $\mM$ (or, equivalently, the heaviest physical
Majorana neutrino mass $m_N$), $\mnu$ and the soft SUSY breaking parameters 
$\mR$ and $\Bnu$. In the literature it is 
often assumed that $\mM$ has a very large value, 
$\mM \sim \cO (10^{14-15}) \gev$, in order to
get $|\mnu| \sim$ 0.1 - 1 eV with 
large Yukawa couplings $ \Ynu \sim \cO(1)$. This is an interesting
possibility since it can lead to important phenomenological implications
due to the large size of the radiative corrections driven by these 
large $ \Ynu$.  We have explored, however, not only these extreme values but 
the full range for $\mM$: $\sim 10^2 - 10^{15} \gev$. 

\begin{figure}[t!]
   \begin{center} 
     \begin{tabular}{cc} \hspace*{-8mm}
  	\psfig{file=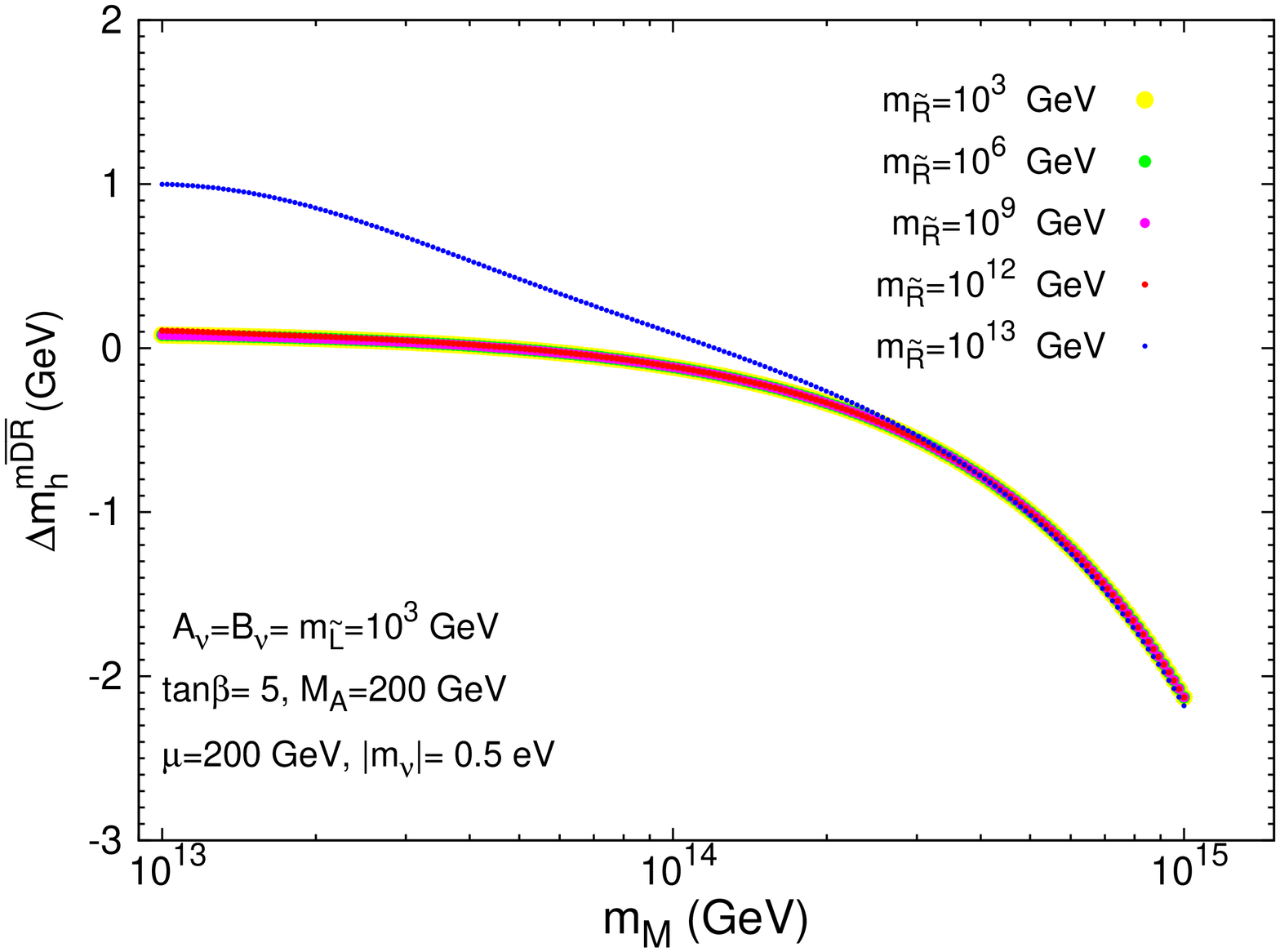,width=50mm,angle=270,clip=} 
	&
        \psfig{file=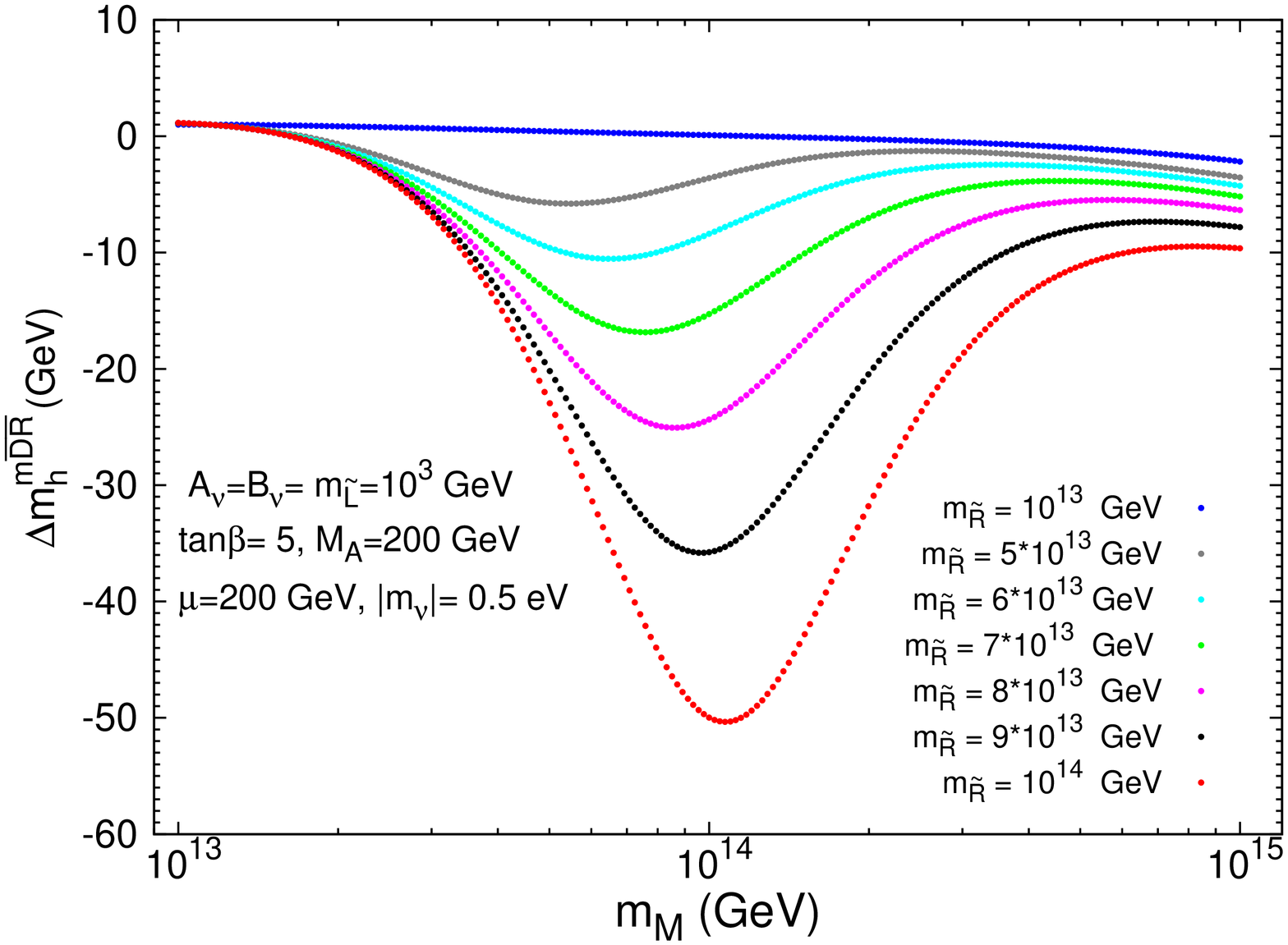,width=50mm,angle=270,clip=} 	  
       \end{tabular}
\caption{One-loop corrections to $\Mh$ from the
neutrino/sneutrino sector as a function of $\mM$  
for:  ${\mR} < 10^{13} \gev$ (left panel) and  
$10^{13}\,\,{\rm GeV}<{\mR}<10^{14}\,\,{\rm GeV}$ (right panel).} 
     \label{fig:mhcorrectionsmR} 
   \end{center}
 \end{figure}
Fig.~\ref{fig:mhcorrectionsmR} shows the predictions for $\Delta m_h^{\mDRbar}$
as a function of $\mM$, for several input $\mR$
values. The Higgs mass corrections are positive and below 0.1 GeV  if
$\mM \lsim 5 \times 10^{13} \gev$ and $\mR< 10^{12} \gev$ (left panel). For 
larger Majorana mass values,  the corrections get
negative and grow up to a few GeV;  $\Delta m_h^{\mDRbar}= -2.15$
GeV for $\mM= 10^{15} \gev$. 
The results in the right plot show that for larger
values of the soft mass, $\mR\gsim 10^{13} \gev$, the Higgs mass
corrections are negative and can be sizeable, a few tens of GeV, reaching their 
maximum values at $\mR \simeq \mM$. 

\begin{figure}[ht!]
   \begin{center} 
     \begin{tabular}{cc} \hspace*{-6mm}
  	\psfig{file=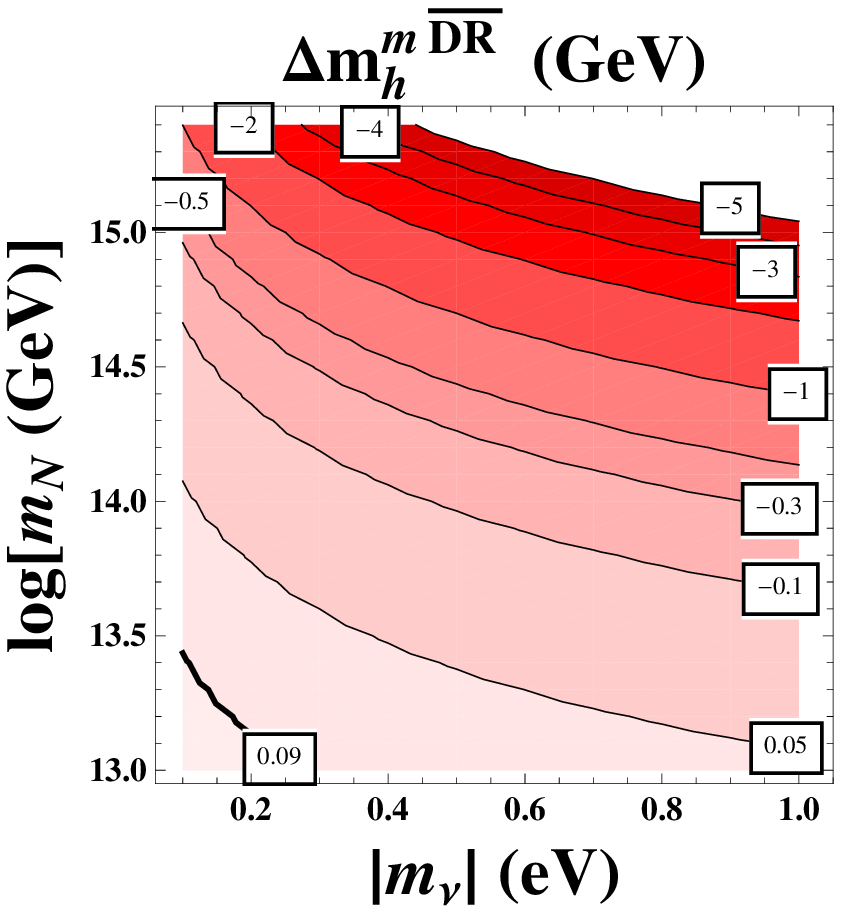,width=69mm,clip=} & 
        \psfig{file=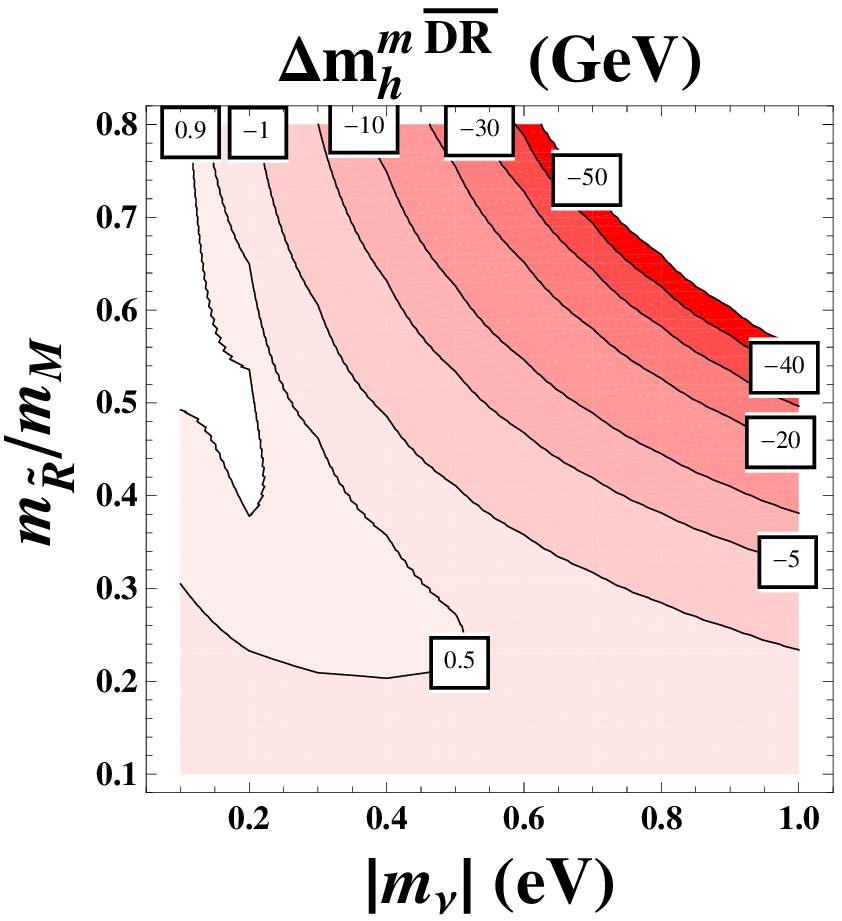,width=69mm,clip=}
       \end{tabular}
     \caption{Contour-lines for the Higgs mass corrections from the
$\nu/\Snu$ sector as a function of: 
$|\mnu|$ and $m_N$ (left panel) and the ratio 
     $\mR/\mM$ and $|\mnu|$ (right panel). The other parameters are: 
$A_\nu=\Bnu=m_{\tilde L}=\mR=
     10^3 \gev, \tb=5, M_A=200 \gev, \mu=200 \gev$.}  
   \label{masscontours1} 
   \end{center}
 \end{figure}

The results of the Higgs mass corrections in terms of the two relevant
physical Majorana neutrino masses, light $|\mnu|$ and heavy $m_N$, 
are summarized in the left plot of Fig.~\ref{masscontours1}. 
 For values of $m_N < 3 \times 10^{13}$ GeV and $|\mnu|< 0.1 -0.3$~eV
 the corrections to $M_h$ are positive and smaller than 0.1 GeV. In this region, the gauge contribution dominates. In fact,
 the wider black  contour line with fixed $\Delta m_h^{\mDRbar}=0.09$
coincides with the prediction for the case where just the gauge part in the
self-energies have been included. This means that 'the distance' of any other
contour-line respect to this one represents the difference in the
radiative corrections respect to the MSSM prediction.
For larger values of  $m_N$ and/or $|\mnu|$ the Yukawa part dominates, and the radiative corrections
become negative and larger in absolute value, up to about $-5 \gev$ 
in the right upper
corner of this figure. These corrections grow in modulus  proportionally to
$\mM$ and $\mnu$, due to the fact 
that the seesaw mechanism impose a relation between the three masses  involved,
$\mD^2 = |m_{\nu}| m_N$.

Finally, we present in the right plot of Fig.~\ref{masscontours1} 
the contour-lines for fixed
$\Delta m_h^{\mDRbar}$ in the less conservative case where $\mR$ is
close to $\mM$. These are displayed as a function of $|\mnu|$ and the ratio
$\mR/\mM$. $\mM$ is fixed to $10^{14}$ GeV. For the interval studied here, 
we see again that the radiative corrections 
can be negative and as large as tens of GeV in the upper right corner of the
plot. For instance, $\Delta m_h^{\mDRbar}=-30 \gev$ for 
$|\mnu|= 0.6$ eV and $\mR/\mM= 0.7$.

\bigskip
To summarize:
for some regions of the MSSM-seesaw parameter space, the corrections to $\Mh$ 
are of the order of several GeV. For all soft SUSY-breaking parameters at 
the TeV scale we find correction of up to $-5 \gev$ to $\Mh$. 
These corrections are substantially larger 
than the anticipated ILC precision of about $50 \mev$ (and also larger than
the anticipated LHC precision of $\sim 200 \mev$). Consequently, they should
be included in any phenomenological analysis of the Higgs sector in the
MSSM-seesaw.

\end{document}